**How effectively does the index of hydrogen deficiency in carbohydrates used in potassium nitrate propellant affect enthalpy change and its performance?**

Sebastian Grabowski



Glossary and Abbreviations

Below is the list of abbreviations used to represent the full name of a chemical, chemical

reaction, data collection method, and symbols in this essay:

**Table 1**

*Chemical compounds abbreviations*

| Full name | Abbreviation |
|---|---|
| Potassium Nitrate | $KNO_3$ or KN |
| D-Sucrose | $C_{12}H_{22}O_{11}$ |
| D-Glucose | $C_6H_{12}O_6$ |
| D-Lactose | $C_{12}H_{22}O_{11}$ |
| D-Galactose | $C_6H_{12}O_6$ |
| D-Sorbitol | $C_6H_{14}O_6$ |
| Xylitol | $C_5H_{12}O_5$ |
| Pentaerythritol | $C_5H_{12}O_4$ or PET |
| Carbon dioxide | $CO_2$ |
| Carbon monoxide | CO |
| Water | $H_2O$ |
| Hydrogen (diatomic) | $H_2$ |
| Nitrogen (diatomic) | $N_2$ |
| Potassium carbonate | $K_2CO_3$ |
| Potassium Hydroxide | KOH |

IHD – Index of hydrogen deficiency

DSC - Differential scanning calorimetry

TGA - Thermogravimetric analysis or thermal gravimetric analysis

$\Delta H$ – change in enthalpy



## Contents









## Introduction

The use of sugar as a fuel in rocket propellant was first mentioned in Captain B.R. Brinley's book "Rocket Manual for Amateurs" in 1960 (see more in Appendix 1), where it was referred to as the "Caramel Candy" propellant. It was listed as one of two widely used amateur propellants, the other being a zinc-sulfur formula referred to as "micrograin." Throughout the years of amateur rocketry and multiple accidents, the Tripoli Rocketry Association decided to ban micrograin on their fields. Because it is so challenging to properly compress the powder to a known value, rocket motors constructed using this formula frequently either have little power and may not lift off or they explode from over pressurization. It is not used in today's world of rocketry anymore. The most popular experimental propellants used by hobbyists around the world are compositions of sugars and potassium nitrate ($KNO_3$). Both professional and amateur rocketeers frequently employ a variety of potassium nitrate ($KNO_3$)-based propellants. Because of its low cost and simple preparation method, it became one of the most widely used rocket fuels. When browsing the internet, one can find numerous recipes for sugar propellants. They vary in terms of the carbohydrate used. This essay's goal is to determine which carbohydrate, when utilized with potassium nitrate as a propellant, has the greatest enthalpy change and, hence, the best performance.

The extended essay investigates the research question: How effectively does the index of hydrogen deficiency in carbohydrates used in potassium nitrate propellant affect enthalpy change and its performance?

To investigate the experimental relationship between Index of Hydrogen Deficiency in carbohydrates used and enthalpy change in combustion of the propellant sample, a calorimeter experiment will be performed in addition to analyzing multiple scientific sources.



Plan of Study

To prove the hypothesis of this work, an index of hydrogen deficiency in various carbohydrates will be used. Sugars (and polyols) used in this work vary in IHD value from 0 to 2.

$$\text{IHD} = \frac{2C + 2 - H}{2}$$

(1)

By using this simple equation, the IHD of any carbohydrate can be simply identified.

**Table 2**

*Thermochemical properties of used carbohydrates*

| Carbohydrate | IHD (calculated) | Enthalpy[1] of combustion (kJ/mol) | Enthalpy of combustion (kJ/g) |
|---|---|---|---|
| D-Sucrose | 2 | -5643.4 | 16.5 |
| D-Glucose | 1 | -2805.0 | 15.6 |
| D-Lactose | 2 | -5667.9 | 16.6 |
| D-Galactose | 1 | -2792.0 | 15.5 |
| D-Fructose | 1 | -2810.4 | 15.6 |
| D-Sorbitol | 0 | -3009.4 | 16.5 |
| Xylitol | 0 | -2564.0 | 16.9 |
| Pentaerythritol | 0 | -2762.0 | 20.3 |

---

[1] Enthalpy of carbohydrates was rounded to one decimal place.



This essay will try to prove that enthalpy change, and energy released by the samples of chemicals are dependent on IHD of carbohydrate used in the mixture. This work will analyze primary source (physical experiment) and secondary sources to support the hypothesis. The work *Fuel-Oxidizer Mixtures: Their Stabilities and Burn Characteristics* analyzes data collected by the author through the use of DSC and TGA equipment. It explores the combustion behaviors of different mixtures of KN with carbohydrates. *Characterization of Potassium Nitrate - Sugar Alcohol Based Solid Rocket Propellants* uses strand burner to provide information about burn rate of some of the chemical mixtures.

This essay will also examine a physical experiment to obtain data on the combustion that is practical in real-world settings. Propellants will be burned in a controlled setting as part of the experiment, and data from the calorimeter will be utilized to analyze the enthalpy of combustion of various mixtures later. These resources ought to offer sufficient data to address the research topic.

### Combustion - Background Information

**Energy**

According to research from Libretexts, the law of conservation of energy refers to an isolated system in which there is no net change in energy and where energy is neither created nor destroyed. Although there is no change in energy, energy can change forms, for example, from potential to kinetic energy. In other words, potential energy (V) and kinetic energy (T) sum to a constant total energy (E) for a specific isolated system. (2020)

$$E = T + V$$

(2)



There is always a corresponding conversion of one kind of energy into another when one substance is transformed into another. The conversion of energy typically involves the emission or absorption of heat, but it can also involve light, electrical energy, or other types of energy.

**Bond Breaking & Bond Forming**

During a chemical reaction, bonds in molecules are broken and other bonds are formed to make different particles. As energy must be added to the system to break a chemical bond, bond breaking is endothermic. On the other hand, the process of forming a chemical bond is exothermic since energy is emitted during this process. For example, the combustion reaction that occurs between carbohydrates and potassium nitrate is an exothermic process; this process, besides emitting heat, also releases energy in the form of light, as evidenced by the flame. The bond energy is the amount of energy required to break one mole of covalent bonds. It is equivalent to the energy required in creating an identical amount of covalent bonds. A covalent bond's strength is evaluated by its bond energy. The more energy that needs to be consumed, the stronger the bond that needs to be destroyed. Similar to this, the more energy is released, the stronger the bond that has been made. Although every molecule has a unique characteristic bond energy, Table 3 shows some generalizations that may be made.



**Table 3**

*Bond energies of different bonds present in carbohydrates*

| Type of bond (single) | Bond energy (kJ/mol) |
|---|---|
| H-H | 432 |
| C-H | 413 |
| C-C | 347 |
| C-O | 358 |
| O-H | 467 |

Heat of reaction = heat absorbed from breaking reactants' bonds – heat released in forming new bonds in products

During a chemical reaction, if the energy required to break bonds in the reactants is greater than the energy released when new bonds are formed in the products, it is an endothermic reaction. If the total energy required to break bonds in the reactants is less than the total energy released when new bonds are formed in the products, it is an exothermic reaction.

**Obtaining sugar alcohols:**

To fully understand behavior and characteristics of polyols, it is needed to know their origin. Methods of obtaining polyols: xylitol, sorbitol, and pentaerythritol are presented below.

Xylitol can be obtained from xylose via fermentation or catalytic hydrogenation. Sorbitol is obtained from glucose, but preferably from fructose, and via fermentation or catalytic hydrogenation. (Galán et al., 2021) Pentaerythritol is a synthetic polyalcohol with the formula C 2OH)4, obtained in the reaction of one acetaldehyde molecule with four molecules of formaldehyde. This reaction takes place in a dilute aqueous solution in the presence of alkaline catalysts. Besides the pentaerythritol a salt of formic acid, most often sodium formate, is the by-product obtained. (Trybula et al., 1990)



**Reactions Mechanisms**

Potassium nitrate decomposes when heated, releasing oxygen, and producing potassium

nitrite ($KNO_2$). That oxygen causes the rapid combustion of sugar (or sugar alcohol).

Chemical structures were created using ChemSketch[2] software.

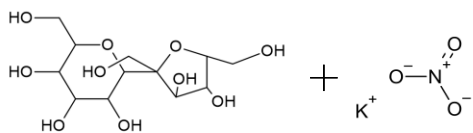

*Fig. 1.* Structure of sucrose reacting with KN.

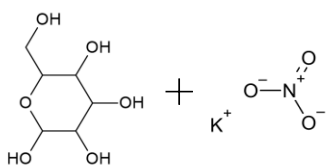

*Fig. 2.* Structure of glucose reacting with KN.

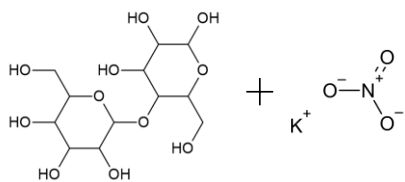

*Fig. 3.* Structure of lactose reacting with KN.

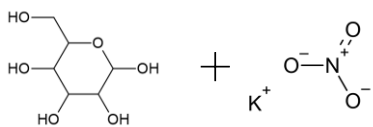

*Fig. 4.* Structure of galactose reacting with KN.

---

[2] ACD/ChemSketch is a molecular modeling program used to create and modify images of
chemical structures.



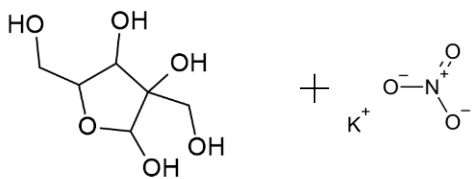

*Fig. 5.* Structure of fructose reacting with KN.

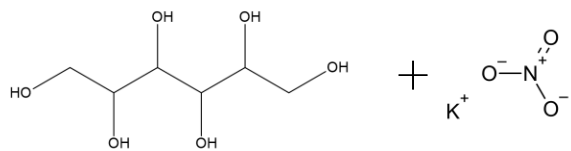

*Fig. 6.* Structure of sorbitol reacting with KN.

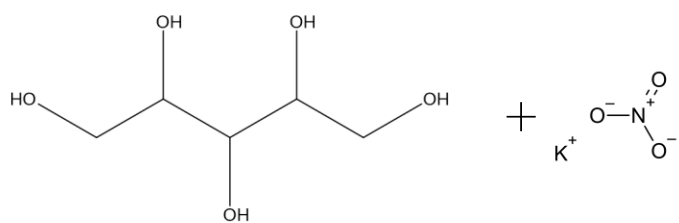

*Fig. 7.* Structure of xylitol reacting with KN.

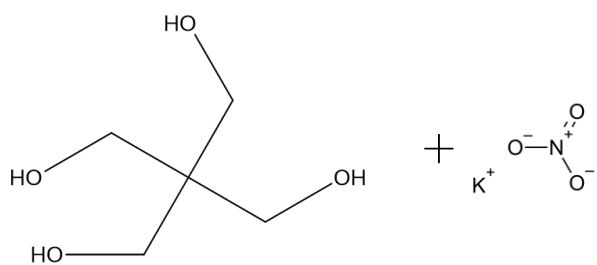

*Fig. 8.* Structure of pentaerythritol reacting with KN.



**Overall Chemical Equations**

A chemical equation can predict how a propellant will burn. The optimal combustion equations of propellants were obtained using PROPEP[3] – chemical combustion simulation program. For carbohydrates used in the analysis with an oxidizer-fuel mass ratio of 65/35, theoretical combustion reactions are:

For sucrose, lactose, and maltose reactions:

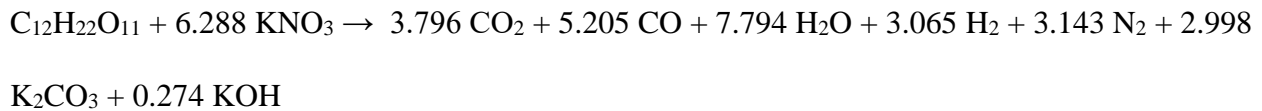

$C_{12}H_{22}O_{11} + 6.288\ KNO_3 \rightarrow 3.796\ CO_2 + 5.205\ CO + 7.794\ H_2O + 3.065\ H_2 + 3.143\ N_2 + 2.998\ K_2CO_3 + 0.274\ KOH$

For glucose, galactose, and fructose reactions:

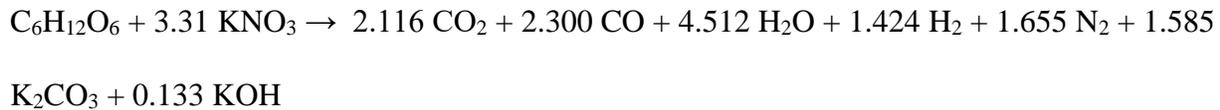

$C_6H_{12}O_6 + 3.31\ KNO_3 \rightarrow 2.116\ CO_2 + 2.300\ CO + 4.512\ H_2O + 1.424\ H_2 + 1.655\ N_2 + 1.585\ K_2CO_3 + 0.133\ KOH$

For sorbitol reaction:

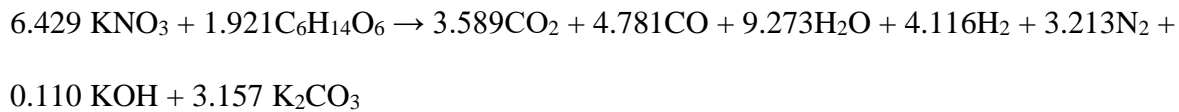

$6.429\ KNO_3 + 1.921\ C_6H_{14}O_6 \rightarrow 3.589\ CO_2 + 4.781\ CO + 9.273\ H_2O + 4.116\ H_2 + 3.213\ N_2 + 0.110\ KOH + 3.157\ K_2CO_3$

For xylitol reaction:

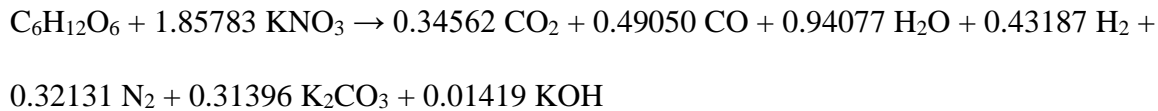

$C_6H_{12}O_6 + 1.85783\ KNO_3 \rightarrow 0.34562\ CO_2 + 0.49050\ CO + 0.94077\ H_2O + 0.43187\ H_2 + 0.32131\ N_2 + 0.31396\ K_2CO_3 + 0.01419\ KOH$

For pentaerythritol reaction:

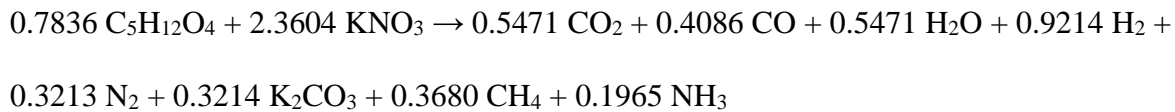

$0.7836\ C_5H_{12}O_4 + 2.3604\ KNO_3 \rightarrow 0.5471\ CO_2 + 0.4086\ CO + 0.5471\ H_2O + 0.9214\ H_2 + 0.3213\ N_2 + 0.3214\ K_2CO_3 + 0.3680\ CH_4 + 0.1965\ NH_3$

---

[3] PROPEP is a program that determines the chemical equilibrium composition for the combustion of a solid or liquid rocket propellant.



Hypothesis

**Enthalpy Change**

It is expected that the enthalpy change would increase as the Index of Hydrogen Deficiency (IHD) rises. It is hypothesized that the more pi bonds/rings an element contains, the more energy will be released in the combustion. Among carbohydrates with the same IHD, the molecular mass would influence enthalpy change character (greater the molecular mass, the greater enthalpy change). The degree of unsaturation increases as the number of pi bonds/rings increases.

**Performance of a propellant.**

It is expected that specific impulse of a propellant will increase as IHD increases. Performance will increase because in the formula containing more pi bonds/rings, the energy from breaking bonds will be higher, therefore enthalpy of combustion should rise.



Argument

**Relationship between carbohydrates with different values of index of hydrogen deficiency**

The writers of the paper *"Fuel-Oxidizer Mixtures: Their Stabilities and Burn Characteristics"* examine energy released, heat flow, and a variety of other combustion-related factors. The analysis of $KNO_3$ combustion with various carbohydrates using a differential scanning calorimeter (DSC) and a thermogravimetric analyzer (TGA) is a component of their study. The propellants created for experimentation in this work had different compositions than the ones used in the listed study. It has been established that composition ratios are very close, and they won't be noticeable when it comes to analysis. The trend of change in enthalpy based on different carbohydrates should be noticeable despite different ratios than in the experiment performed by the author. Stoichiometric weight percentages of mixtures vary between 72.9 and 78.1 (weight % of oxidizer).)

**Characteristics of sugar-based propellants**

Combinations of oxidizers with $KNO_3$ consisted of sucrose, lactose, glucose, and fructose with calculated IHD of 2, 2, 1, and 1 respectively. Oxidizer/fuel ratios used in the experiment were 73.9:26.1 for sucrose and lactose; and 72.9:27.1 for glucose and fructose. Additionally, authors used carbohydrates with IHD = 0: erythritol and pentaerythritol with O/F ratios of 74.9:25.1 and 78.1:21.9 respectively.

**Analysis/Discussion**

Various graphs of DSC analysis, as well as obtained by the authors thermochemical identities of mixtures are shown in their work (Appendix 2). As a control group for carbohydrates, they tested sucrose in various oxidizer combinations. Through their analysis, they provided information that 4:1 KN:sucrose mix releases 1108J/g of energy when combusted.



When studying the combustion behavior of the mixtures, there was no doubt that erythritol used by the authors released the most energy. With 2695J/g at 416°C, this carbohydrate mixture reaction showed a huge difference in enthalpy compared to others.

To help analyze the work, carbohydrates were hierarchized by energy released in the table below:

**Table 4**

*Results of the study – energy released by mixtures of different carbohydrates used with $KNO_3$ as a propellant. Additionally, index of hydrogen deficiency to satisfy analysis of this work.*

| Carbohydrate's Name used in the mixture | Energy Released (J/g) | IHD |
|---|---|---|
| Erythritol | 2695 | 0 |
| Pentaerythritol | 1467 | 0 |
| Sucrose | 1269 | 2 |
| Lactose | 800 | 2 |
| Fructose | 550 | 1 |
| Glucose | 414 | 1 |

These results reject hypothesis of this work and provide a new viewpoint on combustion of those materials. Polyols show a strong lead in the energy released. One of the explanations as to why saturated carbohydrates react to a greater degree than regular sugars is their simpler structure. Pentaerythritol forms crystals, although not as structurally strong as they are in sugar ring form. The crystal behavior of PET would explain why it didn't release as much energy as erythritol. It is simply harder to break by the oxidation reaction with KN. Molecule of PET is also slightly larger than erythritol, having one more carbon atom and two more hydrogen molecules, which increase its molar mass by 14.03g/mol.



In the order of decreasing energy released, after polyols are placed sugars with IHD = 2, followed by carbohydrates with IHD = 1. This behavior could be explained by using ideal chemistry equations (As shown in Overall Chemical Equations section).

Because it has almost twice as big molar mass as glucose, sucrose has the advantage of bonds created by an exothermic reaction. A change in the enthalpy of bonds during the chemical reaction makes all the disaccharides release more energy than monosaccharides. The glycosidic bond, which connects two simple sugar molecules together to form a disaccharide, is mostly unstable and relatively easy to break. This makes sugars with IHD = 2 a better fuel than those with IHD = 1.



**Relationship between carbohydrates with the same value of index of hydrogen deficiency**

Previous works on this topic mostly cover aspects of the burn rate of propellants. To analyze the thesis *"Characterization of Potassium Nitrate - Sugar Alcohol Based Solid Rocket Propellants"* I will look at the results of combustion experiments and their methods of measurement to compare them with my experimental results. This study mainly used a strand burner to determine the burning rate and burn rate related constants for the propellants with IHD = 0: sorbitol, erythritol, and mannitol. This study serves as confirmation that carbohydrates with the same index of hydrogen deficiency don't vary in performance. Based on the hypothesis, saturated sugar alcohols should have the same character of combustion (burn rate). Because all polyols used in the author's work are fully saturated, the difference in combustion can be mostly affected by the molecular weight or crystal geometry of the molecules. To obtain information about the thermochemical properties of the mixtures produced, the burn rate of the propellant will be used.

**Analysis/Discussion**

All propellant mixtures (KN-Sorbitol, KN-Erythritol, and KN-Mannitol) were tested for burn rate at different pressures (Appendix 7). While sorbitol and mannitol formulations are nearly identical, there is some inconsistency with a mix utilizing erythritol as a fuel. As Gudnason (2010) mentions, the burning rates of KN-erythritol are about 20–25% lower than the burning rates of two other mixtures. This situation affected the burn rate coefficient and pressure exponent of the erythritol fuel (Appendix 5).

When it comes to real-world flight applications of all the propellants (performance section of this essay), more thermodynamic equations need to be used to fully answer the research question. The author of the work uses thermochemical and rocketry-specific formulas



that this work will not discuss. Real flight burn characteristics are only possible when the propellant is used in a specific motor. Variables such as specific impulse or exhaust velocity are highly dependent on rocket motor design and the testing environment. Using the proper (and most efficient) carbohydrate propellant does not guarantee the highest impulse and therefore thrust if the engine using that fuel is poorly designed. To keep all the motor characteristics constant, the author used the same motor configuration for all the different propellants and evaluated propellant-specific data (Appendices 5, 6 and 7).

This work will not cover multiple, engineering-type factors that affect the overall performance of the rocket propellant. Because of limitations in both length and the author's knowledge, further investigation of the performance will be based fully on *Characterization of Potassium Nitrate - Sugar Alcohol Based Solid Rocket Propellants.*

After performing MATLAB calculations using the ratio of specific heats for the exhaust gases, the combustion chamber's temperature, the effective molecular weight of the exhaust gases, the pressure at the exit plane of the nozzle, and the pressure inside the chamber, the author came up with the ideal performance estimation for all three propellants working at the pressure of 6.89 MPa (Appendix 4).

Mannitol is an isomer of sorbitol, the only difference between the two is the orientation of the hydroxyl group on carbon 2. Because of very similar chemical characteristics both propellants would have close to identical combustion characteristics. The difference between 164s for sorbitol mixture and 166s for mannitol propellant could be justified by an error/uncertainty.



To analyze and compare all mixtures it is necessary to find their molar amounts used in the experiment. The author of the work used 65/35 O/F ratio in 100 grams batches. Number of moles used in the experiment can be represented by equations on page 12

$$Moles = \frac{grams\ used}{molar\ mass}$$

By using this formula moles used in experiment are as presented below:

Sorbitol: 0.19212823 moles

Mannitol: 0.19212823 moles

Erythritol: 0.28660334 moles

Higher specific impulse of erythritol (167s) can be justified by bigger amount of moles used in the experiment. As seen above, erythritol batch has almost 50% more moles used than other polyols. Due to limitations of potassium nitrate, the change in specific impulse is not so significant as there is not enough oxidizer to support this reaction.

Because of that, erythritol reaction is more likely to release more solid waste than the other mixtures as there might be some remaining chemicals that did not undergo combustion.

All the presented polyols have very similar burn characteristics and therefore almost identical performance. The specific impulses are: 164s for sorbitol mixture, 166s for mannitol mixture, and 167s for erythritol mixture. These values are so close to each other that the actual (not-ideal) real world applications would provide very similar results. This study proved that carbohydrates with the same degree of unsaturation, even if they have different molar masses or structures, are similar (almost identical) in burn characteristics and performance.



## Physical Experiment - Calorimetry

When examining performance and properties, non-ideal conditions should be considered. Simulations by PROPEP are just theoretical. To obtain more information about enthalpy change and its real-world implications, an experiment testing the more realistic enthalpy changes of propellant was performed. The experiment consisted of five trials for each of the mixtures. To keep procedure (Appendix 8) conditions constant, all the trials were run on the same day (in a 5-hour window). Safety and environmental precautions and procedures were presented in Appendix 9.

The pressure in the room was equal to 1019.3 hPa, the temperature was 25 °C, and the humidity of the air was determined to be 44%. As all these factors might influence the chemical reaction and are considered.

**Table 5**

*Experiment's controlled variables*

| Controlled Variables | | |
|---|---|---|
| **Variable** | **Why was it controlled?** | **Method for Control** |
| **Temperature** | Changing temperature of the outside environment could affect results of thermometer and therefore change results of the experiment. | The physical experiment was performed in temperature-regulated lab. |
| **Humidity** | Humidity lowers the temperature of combustion. Water vapor in the air could also react with chemicals in the capsule. | All trials were performed in the same lab, with close to identical conditions regulated by air conditioning system. |



**Results**

To determine energy released by the mixture and the overall ΔH, the following equation was used:

$$\Delta H = Q = mc\Delta T$$

(2)

Where q is the amount of heat gained or lost by a sample, m is mass of water in g, c is specific heat of water (4.18 kJ kG$^{-1}$ °C$^{-1}$), and ΔT is a change in the temperature of the reaction.

To provide an example of how this equation was used, calculations for Galactose 1 trial are presented below:

Knowing that $\Delta T = 3.05$°C and m = 90g:

$$Q = 0.09 kg \times 4.18 \frac{kJ}{kg \times °C} \times 3.05°C$$

$$Q = 1.14741 kJ = 1147.41 J$$

It is worth mentioning that based on Table X, it should be assumed that carbohydrates with IHD=1 should be the ones with the smallest ΔH of the reaction and that theoretically polyols should release the biggest amount of energy.



**Table 5**

*Experimental Results: change in temperature, enthalpy and heat during the chemical reaction.*

| Carbohydrate Trial | ΔT (measured) in degrees Celsius | ΔH (calculated) in Joules | Heat (calculated) released in Joules per gram |
|---|---|---|---|
| Galactose 1 | 3.05 | -1147.41 | 1147.41 |
| Galactose 2 | 2.90 | -1090.98 | 1090.98 |
| Galactose 3 | 3.96 | -1489.75 | 1489.75 |
| Galactose 4 | 3.34 | -1256.51 | 1256.51 |
| Galactose 5 | 3.88 | -1459.66 | 1459.66 |
| Glucose 1 | 5.65 | -2125.53 | 2125.53 |
| Glucose 2 | 5.91 | -2223.34 | 2223.34 |
| Glucose 3 | 5.80 | -2181.96 | 2181.96 |
| Glucose 4 | 5.81 | -2185.72 | 2185.72 |
| Glucose 5 | 6.00 | -2257.20 | 2257.20 |
| Lactose 1 | 5.35 | -2012.67 | 2012.67 |
| Lactose 2 | 5.55 | -2087.91 | 2087.91 |
| Lactose 3 | 5.51 | -2072.86 | 2072.86 |
| Lactose 4 | 5.45 | -2050.29 | 2050.29 |
| Lactose 5 | 5.59 | -2102.96 | 2102.96 |
| Pentaerythritol 1 | 4.58 | -1723.00 | 1723.00 |
| Pentaerythritol 2 | 4.56 | -1715.47 | 1715.47 |
| Pentaerythritol 3 | 4.47 | -1681.61 | 1681.61 |
| Pentaerythritol 4 | 4.63 | -1741.81 | 1741.81 |
| Pentaerythritol 5 | 4.47 | -1681.61 | 1681.61 |
| Sorbitol 1 | 2.42 | -910.40 | 910.40 |
| Sorbitol 2 | 2.43 | -914.17 | 914.17 |
| Sorbitol 3 | 2.64 | -993.17 | 993.17 |
| Sorbitol 4 | 2.51 | -944.26 | 944.26 |
| Sorbitol 5 | 2.68 | -1008.22 | 1008.22 |
| Sucrose 1 | 4.88 | -1835.86 | 1835.86 |
| Sucrose 2 | 4.25 | -1598.85 | 1598.85 |
| Sucrose 3 | 3.77 | -1418.27 | 1418.27 |
| Sucrose 4 | 5.06 | -1903.57 | 1903.57 |
| Sucrose 5 | 3.03 | -1139.89 | 1139.89 |
| Xylitol 1 | 3.49 | -1312.94 | 1312.94 |
| Xylitol 2 | 3.34 | -1256.51 | 1256.51 |
| Xylitol 3 | 3.61 | -1358.08 | 1358.08 |
| Xylitol 4 | 3.64 | -1369.37 | 1369.37 |
| Xylitol 5 | 3.41 | -1282.84 | 1282.84 |

From data collected by the setup (Appendix 8) and presented in Table 5, the highest average change in enthalpy of the reaction was caused by the glucose reaction. The order in which ΔH is decreasing is (carbohydrates used): glucose > lactose > PET > sucrose > xylitol > galactose > sorbitol. Identical order is assigned based on IHD of carbohydrates used in the mixture: 1 > 2 > 0 > 2 > 0 > 1 > 0.



**Discussion**

All carbohydrates have hygroscopic behavior. The polyol with the highest hygroscopicity was determined to be xylitol, followed by sorbitol and PET. Calorimetry testing using polyols turned out to be insufficient. The gelatin capsule used in the study dissolves when in contact with water. When there were no problems with sugars, polyols, when they touched water, this stopped reactions and led to inconsistent data. High hygroscopicity of polyols in comparison to sugars could be explained by a lack of ring structure and less attractive intermolecular forces, allowing bonding with water. Polyols also tend not to form strong crystals because of their high-water solubility.

Results of testing the PET, Xylitol, and Sorbitol are shown in Figure below. There is a noticeable amount of propellant left in the capsule that did not undergo a chemical reaction. The same behavior (to a greater or lesser degree) was present in all the polyols' reactions. Because of that, the calorimeter experiment was not sufficient to provide accurate data about the combustion behavior of KN/polyol mixtures. This error is explained more in the "Random and Systematic Errors" section of this work.

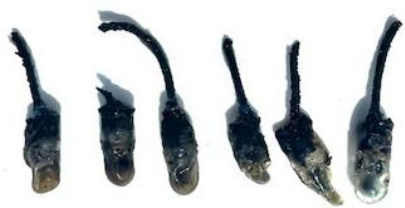

*Fig. 9.* Photo of gelatin capsules after polyols' reaction. From left to right, 2 capsules of: PET, Xylitol, and Sorbitol



Evaluation and Improvements

The dependability of the final observations has several issues that have been found during the evaluation and analysis of the results. This part assesses the investigation's sources' dependability, random and systematic errors, and other factors.

**Random and Systematic Errors**

The most data-influential errors were when testing polyols (as explained in the discussion section). Contact with water caused some of the propellant (in inconsistent amounts) to not react and therefore not contribute to the temperature change. This could be improved by a different preparation procedure for the propellant. Instead of using powdered material, the mixture could be melted to provide maximum bonding of polyol with KN. This would make the propellant react to a greater degree than in its powdered form.

The key systematic error that took place is that not all the heat energy generated by the combustion process is absorbed by the water in the calorimeter. This is because the experiment's basic calorimeter had only minimal insulation. The heat of the reaction was lost to the environment, leading to inaccurate results. This could be resolved by using a differential scanning calorimeter. It could not be suitable for all the propellant mixtures, however, as most apparatus of this type has a maximum operating temperature of around 500 degrees Celsius.

As Oxley et al. (2014) mentioned in the discussion section, "some inconsistency was noticeable when comparing strand burns at the same pressure." Up to 10% of the deviation was recorded. It is believed that the deviation was a result of the non-perpendicular burn of the propellant strand, thus leading to a default in the burn time reading.



Some of the random errors that could occur during the experiment are:

The preparation methods of the mixtures used in the combustion were different. The last difference was the quality of the mixing preparation. A coffee grinder was used in the experiment for this study when the authors of the work used only mortar and pestle, leading to poorer quality propellants in comparison to the coffee grinder used to prepare mixtures for the experiment's purposes.

## Reliability of Sources

The sources included in this paper include books, previous studies, journals, and websites. Some of the scientific work or experimental writings found online are self-published publications that have not yet attained academic recognition or validity. Self-published sources found online are not always trustworthy. The majority of my research's sources are credible since they are well-known academic works published by knowledgeable professionals and databases from widely adopted chemistry websites like Chemistry LibreTexts and PubChem.

A significant issue is that the oxidizer-to-fuel ratios vary among my sources. While secondary sources considered the propellant ratios in stoichiometric mass ratios, the experiment's mixtures were measured in mass ratios. It might lead to inconsistent results that could be dubious for this study to a large degree, as there is not enough oxidizer to supply oxygen to the fuel (carbohydrates). This makes the methods of data collection and analysis insufficient and inadequate to answer the research question stated in this work. Because of that, a primary source in the format of a physical experiment should not be considered a reliable source because it does not show correlations.



Conclusion

"How effectively does the index of hydrogen deficiency in carbohydrates used in potassium

nitrate propellant affect enthalpy change and its performance?"

The sources presented in this work reject the hypothesis. It was proven that IHD is associated

with and has an impact on the enthalpy of combustion of "sugar propellant." The highest

enthalpy change occurs (as expected) in polyol reactions. Carbohydrates with a higher molar

mass and an index of hydrogen deficiency of 2 outperform those with an IHD of 1. *"Fuel-*

*Oxidizer Mixtures: Their Stabilities and Burn Characteristics"* suggests a relationship between

IHD and the heat of combustion. This trend in enthalpy change is shown in Table 4.

It is suggested that there is a correlation between IHD and heat of combustion in *"Fuel-Oxidizer*

*Mixtures: Their Stabilities and Burn Characteristics."* Table 4 displays this enthalpy change

trend. The study "Characterization of Potassium Nitrate-Sugar Alcohol-Based Solid Rocket

Propellants" was chosen to demonstrate that carbohydrates with the same IHD exhibit similar

combustion behavior. This study demonstrates that the performance of carbohydrates with the

same IHD as a propellant is not significantly different. Sorbitol, mannitol, and xylitol have very

comparable combustion properties with nearly identical specific impulses. Physical testing was

done to demonstrate real-world, less-than-ideal propellant performance. However, it did not

substantiate the work's hypothesis. The experiment's findings revealed a significant variation in

the change in enthalpy between trials using the same carbohydrate. The experiment is not a

reliable source of knowledge because of the several restrictions and mistakes stated below. The

environment's loss of heat resulted in a significant amount of ambiguity in the results; hence, this

source should not be used for analysis. The index of hydrogen deficiency may play a less



significant role in choosing a compromised type of carbohydrate due to diverse outcomes that are inconsistent. This prompts exploratory research into the kind of measurement that can be employed in categorizing and evaluating distinct sugars.



Bibliography

1. Abhijeet Singh, D. & Department of Aerospace Engineering MLR Institute of Technology.

(2015). Sugar Based Rocket Propulsion System- Making, Analysis & Limitations.

*International Journal of Engineering Trends and Applications (IJETA)*, *2*(5), 1–8.

http://www.ijetajournal.org/volume-2/issue-5/IJETA-V2I5P7.pdf

2. Delft University of Technology, Olde, M. C., Zandbergen, B. T. C., Jyoti, B. V. S., van den

Wijngaart, J. A. B. J., Kuhnert, F. A., & van Slingerland, J. (2019). Laser Ignition and

Combustion Study of KNO3-Sorbitol based Solid Propellant. *8TH EUROPEAN

CONFERENCE FOR AERONAUTICS AND AEROSPACE SCIENCES (EUCASS)*, 1–11.

https://doi.org/10.13009/EUCASS2019-583

3. Gudnason, M. M. & Department of Chemistry Kemitorvet Technical University of Denmark.

(2010). Characterization of Potassium Nitrate - Sugar Alcohol Based Solid Rocket

Propellants. *Bachelors Thesis*, 1–187.

https://repository.cs.ru.is/svn/mjolnir/research/StrandBurnerMagnusGudnason.pdf

4. Leslie, S., & Yawn, J. (2002). *Proposal for the Inclusion of KNO3/Sugar Propellants to TR*.

Proposal for the Inclusion of KNO3/Sugar Propellants to TRA.

http://www.aeroconsystems.com/motors/sugar_motor/SugarPro_Proposal.pdf

5. Libretexts. (2021a, December 26). *10.4: First Law of Thermodynamics*. Chemistry LibreTexts.

https://chem.libretexts.org/Courses/University_of_Arkansas_Little_Rock/Chem_1300%3

A_Preparatory_Chemistry/Learning_Modules/10%3A_Thermodynamics/10.04%3A_Firs

t_Law_of_Thermodynamics

6. Libretexts. (2021b, December 26). *10.5: Enthalpy of Reaction*. Chemistry LibreTexts.

https://chem.libretexts.org/Courses/University_of_Arkansas_Little_Rock/Chem_1300%3



A_Preparatory_Chemistry/Learning_Modules/10%3A_Thermodynamics/10.05%3A_Ent

halpy_of_Reaction

7. MAE 5540 - Propulsion Systems. Utah State University. (2018). *Chemical Rocket Propellant

    Performance Analysis*. Chemical Rocket Propellant Performance Analysis. http://mae-

    nas.eng.usu.edu/MAE_5540_Web/propulsion_systems/section7/section7.1_2018.pdf

8. Nakka, R. A. (1984). Solid Propellant Rocket Motor Design and Testing. *Solid Propellant

    Rocket Motor Design and Testing*, 1–93.

    https://aeroconsystems.com/tips/RichardNakkaThesis_1984pdf.pdf

9. Olde, M. C. & Delft University of Technology. (2019). Potassium Nitrate Sorbitol Propellant:

    Experimental Investigation of Solid Propellant Characteristics. *Student Theses: Master

    Thesis*, 1–235. https://repository.tudelft.nl/islandora/object/uuid:bd9fbf03-bf45-4bfe-

    aa27-39e1492de3e4?collection=education

10. Oxley, J. C., Smith, J. L., Donnelly, M., & Porter, M. (2014). FUEL-OXIDIZER

    MIXTURES: THEIR STABILITIES AND BURN CHARACTERISTICS. *International

    Journal of Energetic Materials and Chemical Propulsion*, *13*(6), 517–557.

    https://doi.org/10.1615/intjenergeticmaterialschemprop.2014011485

11. *Recrystallized Rocketry*. (2016). Recrystallized Rocketry. https://www.jamesyawn.net/

12. *Richard Nakka's Experimental Rocketry Site*. (2022). Richard Nakka's Experimental

    Rocketry Web Site. https://www.nakka-rocketry.net

13. *Rocket Thermodynamics*. (2012). Rocket & Space Technology.

    http://www.braeunig.us/space/thermo.htm#enthalpy

14. *Sugar Fuels*. (2022). Sugar Fuels. http://www.ajolleyplace.com/fuel.html



15. Vyverman, Antoon (Tony) and Youth & Space 1978. *The Potassium Nitrate Sugar Propellant*. Working Group Propulsion, 1978, www.vro.be

16. Pagonis, D., Claflin, M. S., Levin, E. J. T., Petters, M. D., Ziemann, P. J., & Kreidenweis, S. M. (2017, June 30). Hygroscopicity of Organic Compounds as a Function of Carbon Chain Length and Carboxyl, Hydroperoxy, and Carbonyl Functional Groups. *The Journal of Physical Chemistry A*, *121*(27), 5164–5174. https://doi.org/10.1021/acs.jpca.7b04114

17. Galán, G., Martín, M., & Grossmann, I. E. (2021). Integrated Renewable Production of Sorbitol and Xylitol from Switchgrass. *Industrial &Amp; Engineering Chemistry Research*, *60*(15), 5558–5573. https://doi.org/10.1021/acs.iecr.1c00397

18. Trybula, S., & Terelak, K. (1990). A Polish process for pentaerythritol manufacture. *Pigment &Amp; Resin Technology*, *19*(5), 7–8. https://doi.org/10.1108/eb042719

19. *NIST Chemistry WebBook*. (2021, August 12). https://webbook.nist.gov/chemistry/

20. *Amateur Rocketry Motor (Engine) Propellant*. (n.d.). Retrieved October 25, 2022, from https://jacobsrocketry.com/aer/propellant.htm



Appendices

**Appendix 1**

*Briefly History of Sugar Propellants adapted from Proposal for the Inclusion of KNO3/Sugar*

*Propellants to TRA (written by Stuart Leslie and James Yawn)*

**The following is a general timeline illustrating the development of SPs over the**

**years**.

**1944** First experiments with KN/sucrose by William Colburn of the Rocket Missile Research

Society (later Rocket Motor Research Group) followed by first launch of a sugar rocket in 1947.

The first propellant, designated TF-1 was dry-mixed KN/sucrose moistened with water and

pressed into the motor tube. (From Bill Colburn via Richard Nakka′s website)

**1950** Melting of KN/sucrose by Dirk Thysse allows casting of propellant grains, thus different

configurations are available and larger motors can be produced. (Colburn via Nakka)

**1957** First use of hydraulic press to make grains by Bill Colburn allows creation of Bates-type

grains without the need to melt propellant components.

**1960** Publication of ″Rocket Manual for Amateurs″ by B. R. Brinley includes a brief description

of KN/sucrose propellant. This book has inspired many amateurs (notably Richard Nakka) to try

sugar propellants, as well as serving as a manual for many high-school rocketry clubs, which

often favored sugar propellant (John Wickman, personal correspondence.)

**1975** BVRO (Flemish Rocketry Organization) begins first scientific investigation of this

KN/sucrose propellant and its performance characteristics. Culminates in the

**1980** publication of ″The Potassium Nitrate - Sugar Propellant″ by Antoon Vyverman. This

group recognized problems of case-bonding in large sugar motors. Contains first mention of



KN/sorbitol and KN/mannitol, first used by this group in 1977. (Vyverman 1980, Vyverman, personal correspondence)

**1983** David Sleeter of Teleflite Corporation, releases the book ″Building Your own Rocket Motors, ″ and a booklet entitled: ″The Incredible Five-Cent Sugar Rocket″ using KN/sucrose and sulfur. These publications were advertised in magazines such as Popular Science, and so presumably found wide audience.

**1993** NERO performs substantial tests of KN/sorbitol, which allows melting and casting at lower temperature than sucrose, perceived to be safer. (Vyverman, personal correspondence)

**1996** First use of KN/dextrose by Richard Nakka. KN/DX offers a melting point only a little higher than KN/sorbitol, with a more predictable burn rate. Also, rocket discussion group opens, facilitating discussion of amateur/experimental rocketry in general. Sugar propellants are a common topic on this list.

**1997** Richard Nakka opens his website on experimental rocketry, devoted primarily to the use of sugar propellants. His creative and careful technical work provides a scientific foundation for experimenters working with sugar propellants. It includes substantial treatments of rocketry theory in general, so as to be commendable to users of other propellants as well. Also, Al Bradley uses hydraulic press to compress sugar propellant moistened with 60/40 water/ethanol, resulting in grains which air-dry to a very hard consistency. Proposal for the Inclusion of KNO3/Sugar Propellants to TRA, October 4, 2002, Page 4

**1998** Publication of Rocket Boys by Homer Hickam and subsequent movie October Sky (1999) stirs new interest in rocketry, creating a distinct class of ″Born-Again Rocketeers.″ KN/sucrose was one of the propellants used by the ″Rocket Boys,″ ostensibly co-invented by them independent of knowledge of previous experimenters. (Hickam 1998)



**2000** Invention of the ″Candymatic,″ by Paul Kelly allows automated, remote melting of propellants. It is a bread machine modified to heat and stir propellant to the melting point, allowing the operator to remain at a distance. (2001 Jay Ward places a photo of the Candymatic on the Web.)

**2001** Publication of the recrystallization process by Jimmy Yawn resolves some limitations of KN/sucrose.

**2002** Substantial dialogue regarding sugar propellants on the Arocket discussion list prompts formation of the SugPro list, specifically for the discussion of sugar-based propellants. List owners are Dave McCue and David Muesing. SugPro currently has 100 members (September 2002).

**2002** ″Yuv″ reports melting of KN/sorbitol in a boiling-water bath. The propellant mix is enclosed in a plastic bag and immersed in hot water. This may be the safest method yet of melting sugar propellants.

**2002** Two different companies produce and market kits for making sugar-propelled rocket engines. Woody Stanford of Stanford Systems and Jon Drayna of October Science each produce such kits. Sugar propellant knocks at the door of consumer rocketry.



**Appendix 2**

*Results of differential scanning calorimeter (DSC) and Thermogravimetric Analyzer (TGA)*

*adapted from Fuel-Oxidizer Mixtures: Their Stabilities and Burn Characteristics. Temperature*

*endotherms & exotherms for DSC & SDT at 20◦/min (heat release J/g)*

| | $KIO_3$ | | $KNO_3$ | | $NH_4NO_3$ | | $KNO_2$ | |
|---|---|---|---|---|---|---|---|---|
| | Endotherm | Exotherm | Endotherm | Exotherm | Endotherm | Exotherm | Endotherm | Exotherm |
| Phase change | nothing in DSC | | 132(53), 222(3) | | 55(22), 130(40) | | 45(13) | |
| Melt | 553d(609) | | 330(29) | | 167 | | 424(82) | |
| Melt KX | 680(75) | | | | | | | |
| Decompose | 831-**900**(275) | | 702(433) | | | 261-**316**-336 (1407) | 517(2), 903(17) | 930(176) |
| Sucrose | 62 152(19), 678(8) | 62 156-**187** (1736), 438(137) sdt | 18 137(22), 181(73), 219(14) | 18 219-**265** (298), 359-**393** (400 or 1269) | 22 54(16), 127(15) | 22 147-**176** (2627), 296(513) | 55 192(57) | 55 212-**239**-290(1618) |
| Sucrose 20% | 77 160(39) | 77 163-**182**(939) dsc, **434**(207) sdt | 72 131(45), 173(20) | 72 **381**-**396**-413-**453**(1108) | 68 56(14), 129(33) | 68 151-**176**, **202**(2084), 283-**295**-333(1010) | 78 172(41), 940(21) | 78 177-**212**-222-251(1777), 323-**333**(65) |
| Lactose | | | 19 132(9), 151(70), 212(9) | 19 **271**(300), 379-**394**(800) | 23 55(10), 107(110) | 23 160-**182**-209, 221 (1493), 311(183) | | |
| Fructose | 63 123(27), 678(10) | 63 128-156 (1491), 430(117) | 21 122(72) | 21 230-314-380 (244), **409**(550) | 25a 55(15), 130(80) | 25a 150-**170**(3475), 336(607) | | 56 136-**170**-218 (944) |
| Glucose | | | 20 133-152-165 (140) | 20 **271**-305 (370), **392**(414) | 24 54(11), 107(90) | 24 153-**188**(2400), 309(318) | | |
| Pentaerythritol | | | (23), 190(139), 354(164), 687(849) | 440, **470**(1467), 765, 842, 940 | 33 54(13), 126 (38), 175(9) | 33 242-**265**(2008) | | |
| Erythritol | 64 123(119), 608(2), 668(15) | 64 178(931) | 34 113(125), 130(32) | 34 298(30), **416**(2695) | 35 86(59), 119(11) | 35 255-**261**(1653) | 57 116(43) | 57 250-**318**-357 (1041) |
| Surfur | 88 114, 119(28) | 88 270-**363**-400(2411) | 40a 115, 119, 131(65), 189(12) | 40a 300-**327**-420(1092) | 39 116(40) | 39 198-**217**(2379) | | 87 251-**294**-310(1962) |
| Charcoal | | | 43 129(22), 332(47) | 43 **461**(2182) | | 41 **221**(1472) | | |
| | $KIO_3$ | | $KNO_3$ | | $NH_4NO_3$ | | $KNO_2$ | |



## Appendix 3

*Oxidizers with 50wt% fuel DSC response & average J/g (number runs) adapted from Fuel-Oxidizer Mixtures: Their Stabilities and Burn Characteristics.*

| | $KIO_4$ | °C | $KIO_3$ | °C | $KBrO_3$ | °C | $KClO_3$ | °C | $KClO_4$ | °C | $NH_4ClO_4$ | °C | $KNO_3$ | °C | $KNO_2$ | °C | $NH_4NO_3$ | °C | $KMnO_4$ | °C | $KCr_2O_7$ |
|---|---|---|---|---|---|---|---|---|---|---|---|---|---|---|---|---|---|---|---|---|---|
| Thermal change | 350 | °C | 553 | °C | 417, 460 | °C | 350, 415 | °C | 305 | °C | 248, 400 | °C | 133, 330 | °C | 323, 430 | °C | 130, 167, 316 | °C | 260, 305 | °C | 398 |
| Sugars (130–190°C) | melt | | melt | | melt | | melt+ | | >400 | | melt+ >248 | | 262, >330 | | melt | | melt AN & fuel | | melt+ | | >398 |
| Peak temperature °C | 167 | | 187 | | 206 | | 175 | | 495 | sdt | 338 | | 386 | | 254 | | 171 | | 239 | | 402 |
| J/g for 50 wt% sucrose | 2054 | 2 | 1643 | 2 | 1110 | 6 | 2033 | 3 | 1320 | 2 | 2341 | 2 | 926 | 3 | 1231 | 3 | 2136 | 4 | 1169 | 3 | 50 |
| PE (190, 233, 305) | – | | – | | – | | melt+ | | – | | – | | >400 | | – | | 269+ | | – | | – |
| Peak temperature °C | | | | | | | 270 | | | | | | 471 | | | | 267 | | | | |
| Heat released J/g | | | | | | | 1797 | 3 | | | | | 1669 | 4 | | | 2087 | 4 | | | |
| Erythritol (124, 330) | melt+ | | melt | | melt+ | | 200 | | >305+ | | >248+ >400 | | >400 | | 270 | | 270 | | melt+ | | >398 |
| Peak temperature °C | 142 | sdt | 185 | | 229 | | 253 | | 181 | | 313 | | 413 | | 316 | 5 | 269 | | 325 | 2 | 390 |
| Heat released J/g | 1327 | 2 | 871 | 3 | 1160 | 3 | 2314 | 2 | 881 | | 3616 | 3 | 2471 | 3 | 1014 | 5 | 1817 | 4 | 1836 | 2 | 129 |
| Sulfur (116, 180, 315) | 250 | | 220 | | – | | 150 | | >400 | | >248 | | ~300 | | 290 | | 170 | | 270 | | – |
| Peak temperature °C | 303 | | 298 | | | | 194 | | 468 | | 422 | | 333 | | 299 | | 219 | | 309 | | |
| Heat released J/g | 1410 | 4 | 659 | 4 | | | 1031 | 3 | 1612 | 3 | 1747 | 3 | 916 | 4 | 1006 | 3 | 2103 | 2 | 808 | 3 | |
| Charcoal | – | | – | | – | | 335 | | >400 | | >248 | | >400 | | – | | 223 | | – | | >398 |
| Peak temperature °C | | | | | | | | | 526 | sdt | 450 | | 467 | | | | | | | | |
| Heat released J/g | | | | | | | 1470 | 2 | 1482 | 4 | 951 | 3 | 1403 | 4 | | | 1611 | 2 | | | 100 |

fuel controlling   oxidizer contolling   fuel or oxidizer controlled



**Appendix 4**

*Ideal performance estimation of KN-Sorbitol, KN-Erythritol and KN-Mannitol at 6.89 MPa.*

|  | KN-Sorbitol | KN-Erythritol | KN-Mannitol |  |
|---|---|---|---|---|
| $T_c$ | 1600 | 1608 | 1637 | K |
| $\gamma_{gas}$ | 1.2391 | 1.2416 | 1.2376 |  |
| $\gamma_{mix}$ | 1.1361 | 1.1390 | 1.1362 |  |
| $\gamma_{2ph}$ | 1.0423 | 1.0425 | 1.0427 |  |
| $v_e$ | 1609 | 1639 | 1628 | m/s |
| $I_{sp}$ | 164 | 167 | 166 | s |
| c* | 908 | 925 | 920 | m/s |

**Appendix 5**

*Burn rate coefficient (a) and pressure exponent (n) for KN-Sorbitol, KN-Erythritol and KN-Mannitol propellant*

|  | a | n |
|---|---|---|
|  | mm/sec/MPa |  |
| KN-Sorbitol | 5.132 | 0.222 |
| KN-Erythritol | 2.903 | 0.395 |
| KN-Mannitol | 5.126 | 0.224 |



**Appendix 6**

*Comparison of experimental results for KN-Sorbitol, KN-Erythritol and KN-Mannitol*

*propellants.*

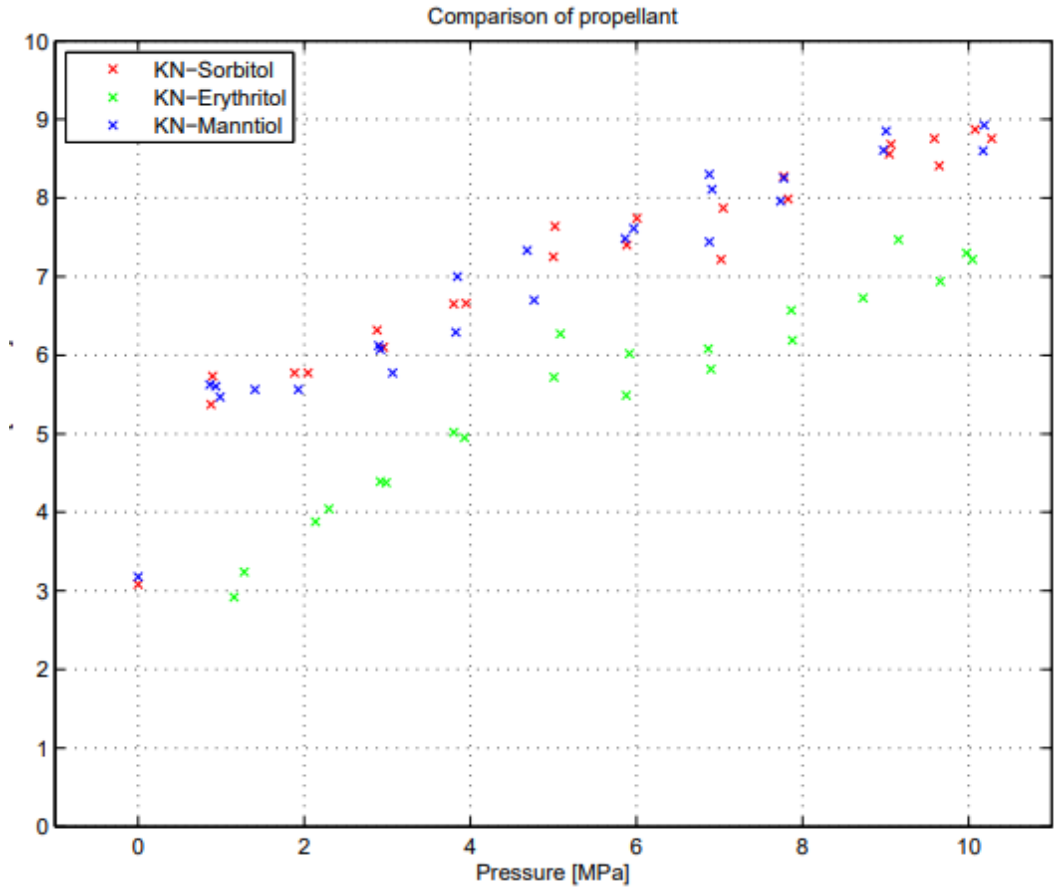



**Appendix 7**

*Experimental results for KN-Sorbitol, KN-Erythritol and KN-Mannitol propellants. Vessel pressure is shown as gauge and absolute pressures. (From Characterization of Potassium Nitrate - Sugar Alcohol Based Solid Rocket Propellants)*

| KN-Sorbitol | | | KN-Erythritol | | | KN-Mannitol | | |
|---|---|---|---|---|---|---|---|---|
| Vessel Pressure | | Burn Rate | Vessel Pressure | | Burn Rate | Vessel Pressure | | Burn Rate |
| MPag | MPaa | mm/sec | MPag | MPaa | mm/sec | MPag | MPaa | mm/sec |
| 0 | 0.10 | 3.08 | 0 | 0.10 | NA | 0 | 0.10 | 3.18 |
| 0.88 | 0.98 | 5.37 | 1.16 | 1.26 | 2.92 | 0.87 | 0.97 | 5.62 |
| 0.90 | 1.00 | 5.73 | 1.28 | 1.38 | 3.24 | 0.94 | 1.04 | 5.60 |
| 1.89 | 1.99 | 5.77 | 2.14 | 2.24 | 3.88 | 0.99 | 1.09 | 5.47 |
| 2.05 | 2.15 | 5.77 | 2.30 | 2.40 | 4.04 | 1.41 | 1.51 | 5.56 |
| 2.88 | 2.98 | 6.32 | 2.92 | 3.02 | 4.39 | 1.93 | 2.03 | 5.56 |
| 2.96 | 3.06 | 6.10 | 2.99 | 3.09 | 4.38 | 2.07 | 2.17 | 5.77 |
| 3.80 | 3.90 | 6.65 | 3.80 | 3.90 | 5.02 | 2.92 | 3.02 | 6.07 |
| 3.95 | 4.05 | 6.66 | 3.93 | 4.03 | 4.95 | 2.99 | 3.09 | 6.12 |
| 5.00 | 5.10 | 7.25 | 5.01 | 5.11 | 5.72 | 3.83 | 3.93 | 6.29 |
| 5.02 | 5.12 | 7.64 | 5.09 | 5.19 | 6.27 | 3.85 | 3.95 | 7.00 |
| 5.89 | 5.99 | 7.40 | 5.88 | 5.98 | 5.49 | 4.69 | 4.79 | 7.33 |
| 6.01 | 6.11 | 7.74 | 5.92 | 6.02 | 6.02 | 4.77 | 4.87 | 6.70 |
| 7.02 | 7.12 | 7.22 | 6.87 | 6.97 | 6.08 | 5.87 | 5.97 | 7.48 |
| 7.05 | 7.15 | 7.87 | 6.90 | 7.00 | 5.82 | 5.97 | 6.07 | 7.61 |
| 7.78 | 7.88 | 8.27 | 7.87 | 7.97 | 6.57 | 6.88 | 6.98 | 7.44 |
| 7.83 | 7.93 | 7.99 | 7.88 | 7.98 | 6.19 | 6.88 | 6.98 | 8.30 |
| 9.05 | 9.15 | 8.56 | 8.73 | 7.83 | 6.73 | 6.91 | 7.01 | 8.11 |
| 9.07 | 9.17 | 8.68 | 9.66 | 9.76 | 6.94 | 7.74 | 7.84 | 7.96 |
| 9.59 | 9.69 | 8.76 | 9.16 | 9.26 | 7.47 | 7.78 | 7.88 | 8.25 |
| 9.65 | 9.75 | 8.41 | 9.98 | 10.08 | 7.30 | 8.98 | 9.08 | 8.61 |
| 10.08 | 10.18 | 8.87 | 10.05 | 10.15 | 7.22 | 9.01 | 9.11 | 8.85 |
| 10.28 | 10.38 | 8.76 | | | | 10.18 | 10.28 | 8.60 |
| | | | | | | 10.19 | 10.29 | 8.93 |



**Appendix 8**

*Method of data collection in the experiment*

Chemicals (KNO3 and carbohydrates) are weighted using digital scale (uncertainty ± 0.0005g). They're well mixed with each other and grinded to a fine texture using coffee grinder. The propellant is heated (and melted) during the common and well-known casting procedure. However, due to the fire risk and the lack of the necessary hardware to maintain the heating plate at a constant temperature, this method of preparation was not adopted. A custom "casting" method consisted of inserting 1.0000 gram (+/- 0.0005g) of prepared earlier propellant into gelatin capsules (medicine capsules). After that, a hole was bored in the capsule and a pyrotechnic fuse was inserted there to make contact with the propellant. Each capsule was approximately 0.1250 grams (+/- 0.0005g) each (mass varied between 0.1220 to 0.1265g). Energy released by the capsule itself was approximated at 0.2375J (based on the information that pure gelatin releases 1.9kJ/Kg). This amount of energy is negligible when it comes to enthalpy change in the calorimeter during combustion of the propellant. The fuse was glued to the capsule, so it does not separate while burning. Glue also protects chemicals from reacting with the outside environment. Enthalpy of the fuse was not included in calculations as well because it was identical (or close to identical) for all the trials.

To obtain real measurements of enthalpy and character of combustion of chosen propellants, standard calorimetry was used as a data collection method. Previously prepared capsules with propellants inside were placed in the aluminum calorimeter with 90g of pure water. It was decided to not use bomb calorimeter as it could potentially over-pressurize and rupture during combustion. Calorimeter setup consisted of double walled aluminum calorimeter



with Styrofoam liner, Vernier LabQuest interface with a temperature sensor, and a laptop with Vernier Logger Lite software.

Data collection procedures were:

1. Software (Logger Lite) was running for 80-180sec. to record the whole process of the reaction including the decrease of temperature after the reaction has occurred.

2. The capsule with the propellant was inserted into the calorimeter containing 90 grams of pure water.

3. A fuse connected to the propellant was lighted up and the thermometer sensor started recording data.

4. After the reaction took place, thermometer data was transferred to software and saved for later analysis. The calorimeter was cleaned and dried before next trial.

Data collection consisted of 5 trials for each of the propellant mixtures.

Data collection setup:

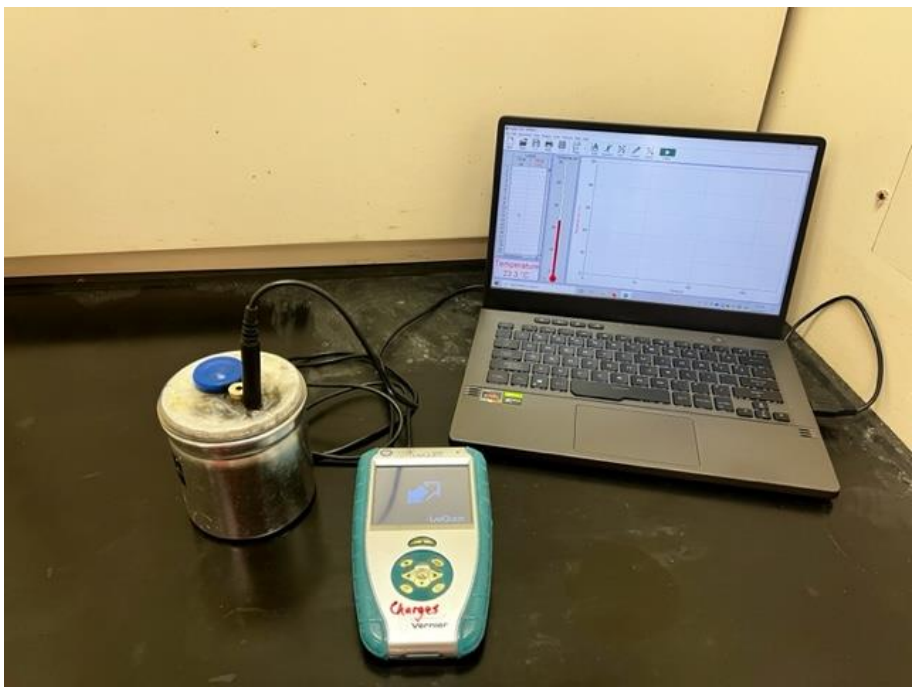



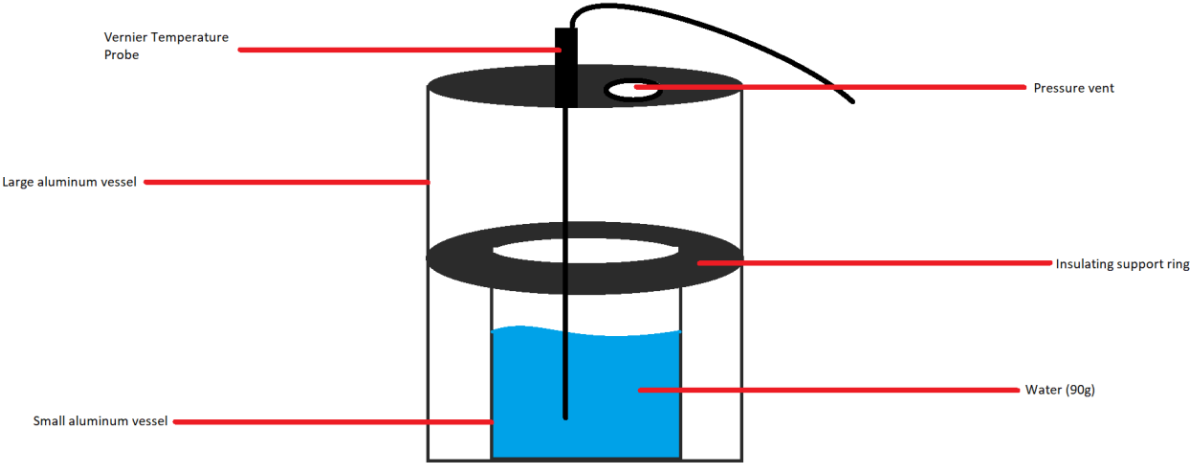



## Appendix 9

*Safety and environmental precautions and procedures for the physical experiment.*

### Safety precautions and procedures

Some safety precautions have been taken because some of the compounds are hazardous. Potassium nitrate and carbohydrates had to be stored in separate containers, ideally in distinct locations, before any substances were mixed to make propellant combinations. Both sealed goggles and nitrile gloves were always worn when making gelatin capsules and combining ingredients. A N95 mask was also worn because the molecular size after being ground in the coffee grinder was so small. These safety measures result from potassium nitrate's toxicity to the skin and respiratory system (as seen in Appendix 11).

Carbon dioxide, carbon monoxide, water, hydrogen, nitrogen, potassium carbonate, and potassium hydroxide are all made when different types of carbohydrates are burned.

Carbon monoxide is a tasteless, odorless, and clear gas. Carbon monoxide poisoning is caused by inhaling combustion fumes. In extreme cases, CO poisoning may cause loss of consciousness or death. The physical experiment conducted for this study releases small amounts of CO, probably harmless to humans. Despite that, safety precautions in terms of using laboratory fume hood were taken.

The release of hydrogen gas throughout the process also supported the use of the fume hood. Although non-toxic, hydrogen gas is very flammable. It can be ignited by sparks, heat, or static electricity. The experimental equipment was located away from all sources of heat and electric sparks to ensure a secure research environment.



Potassium carbonate is harmful when in contact with skin. It is also poisonous when swallowed or inhaled. To ensure safety when running experiment trials, N95 masks, nitrile gloves, and sealed goggles were always worn.

**Environmental precautions and procedures**

Carbon dioxide and water vapor are two greenhouse gases produced by this experiment. However, because these chemicals are present in such small quantities, their environmental impact is minimal. However, potassium carbonate is hazardous to the environment and shouldn't be discarded as waste. This chemical required a reaction with hydrochloric acid in order to be neutralized.

According to the balanced chemical equation below, neutralization of K2CO3 results in the production of water, carbon dioxide, and nontoxic potassium chloride.

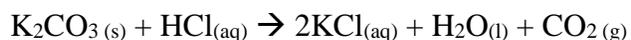

$K_2CO_3\text{ (s)} + HCl_{(aq)} \rightarrow 2KCl_{(aq)} + H_2O_{(l)} + CO_2\text{ (g)}$

Equimolar ratios of potassium carbonate and hydrochloric acid were necessary for the reaction. On page 14 of the report, under the heading "Overall Chemical Equations," are chemical equations that were used to predict how much potassium carbonate is produced during combustion reactions.